\documentclass[%
 reprint,
superscriptaddress,
 amsmath,amssymb,
 aps,prapplied,floatfix
]{revtex4-2}
\usepackage{colortbl}
\usepackage{color}
\usepackage{graphicx}% Include figure files
\usepackage{dcolumn}% Align table columns on decimal point
\usepackage{bm}% bold math
\usepackage{hyperref}% add hypertext capabilities

\usepackage{xcolor}
\colorlet{RED}{red}

\usepackage{amsmath}
\usepackage{braket}
\usepackage[normalem]{ulem}

\hypersetup{pdfstartview={FitH},pdfpagemode={UseNone}, colorlinks=true,bookmarks=false,allcolors=blue, breaklinks=true, bookmarksopen=true, pdfnewwindow=true}
\usepackage[all]{hypcap}

\graphicspath{{pict/}{figures/}{}}

\usepackage[english]{babel}

\begin{document}

\title{Complex structure and characterization\\of multi-photon split states in integrated circuits}
%\title{Complex structure and direct characterization of multi-photon split states}
%\title{Coupled waveguide neural network for multi-photon split state tomography}

\author{Jihua Zhang}
\email{jhzhanghust@gmail.com}

%,\authormark{1,*} and 
\author{Andrey A. Sukhorukov}
\email{andrey.sukhorukov@anu.edu.au}
%,\authormark{1,\#}}

\affiliation{ARC Centre of Excellence for Transformative Meta-Optical Systems (TMOS), Department of Electronic Materials Engineering, Research School of Physics, The Australian National University, Canberra, ACT 2601, Australia}
%\authormark{2}ARC Centre of Excellence for Transformative Meta-Optical Systems (TMOS), Australia}
% \authormark{3}Currently with the Department of Electronic Journals, The Optical Society (OSA), 2010 Massachusetts Avenue NW, Washington, DC 20036, USA}

%\expandafter
%\email{\authormark{*}\emailLink{jhzhanghust@gmail.com}\\ \authormark{\#}\emailLink{andrey.sukhorukov@anu.edu.au}} %% email address is required



\date{\today}% It is always \today, today,

\begin{abstract}
Multi-photon split states, where each photon is in a different spatial mode, represent an essential resource for various quantum applications, yet their efficient characterization remains an open problem.
%\textcolor{red}{How to efficiently characterize the multi-photon indistinguishability of such states with arbitrary form of density matrix remains an open question.} 
Here, we formulate the general structure of their reduced spatial density matrices 
%\sout{based on the symmetry analysis} \textcolor{red}{after integrating over the photon frequencies}, 
and identify the number of real and complex-valued independent coefficients, which in particular completely determine the distinguishability of all photons.
%and any subset of photons
Then, we show that this density matrix can be fully characterized by measuring correlations after photon interference in a static integrated circuit, where the required outputs scale sub-quadratically versus the number of photons. We present optimized circuit designs composed of segmented coupled waveguides, representing a linear optical neural network, which minimize the reconstruction error and facilitate robustness to fabrication deviations. 
\end{abstract}

\maketitle

\section{Introduction}

The multi-photon split states (M-PSS), where each photon is in a different spatial mode of free-space beams, fibers, or waveguides, represent an essential resource for fundamental tests of quantum mechanics and various applications in quantum simulations and computations~\cite{Nielsen:2011:QuantumComputation}. For example, many photon interference and distinguishability experiments are based on M-PSSs where each photon is injected from a different spatial port~\cite{Hong:1987-2044:PRL, Bouchard:2021-12402:RPP, Menssen:2017-153603:PRL, Spagnolo:2013-1606:NCOM, Brod:2019-63602:PRL, Pont:2022-31033:PRX, Giordani:2020-43001:NJP}. 
%In addition, the most commonly used quantum photonic source via \textcolor{red}{noncollinear} spontaneous parametric down conversion is a two-photon split state where two entangled photons are \textcolor{red}{generated to different directions} \sout{or polarization modes}~\cite{Kwiat:1995-4337:PRL}. 
%\sout{In the early experiments on photonic boson sampling, the initial sources were all based on such a split state} 
%\textcolor{red}{
Furthermore, photon boson sampling experiments that can demonstrating quantum advantage commonly employ PSS sources that provide multiple indistinguishable photons with each of the photons injected to a different port of a linear optical network~\cite{Spring:2013-798:SCI, Broome:2013-794:SCI, Tillmann:2013-540:NPHOT, Crespi:2013-545:NPHOT, Shchesnovich:2015-63842:PRA, Renema:2018-220502:PRL}. Remarkably, PSSs with more than two photons were recently shown to possess a multi-photon collective phase beyond the real-valued pair-wise photon distinguishability measure, opening new degrees of freedom for quantum information~\cite{Menssen:2017-153603:PRL, Shchesnovich:2018-33805:PRA, Jones:2020-123603:PRL}. 
Therefore, the characterisation of PSSs is of great importance from fundamental and practical perspectives.

%The resource-efficient characterization of M-PSSs remains an 

Importantly, the indistinguishability of all photons in M-PSS cannot be inferred from the  distinguishability of constituent photon pairs, which demands the characterization based on multi-photon interference. Efficient protocols for witnessing multiphoton indistinguishability were developed and realized with reconfigurable multi-port interfrometers~\cite{Pont:2022-31033:PRX} under the assumption of a specific density matrix form, whereas this remains an open problem for general states (see a discussion in the Supplementary of Ref.~\cite{Brod:2019-63602:PRL}).
%is fundamentally different from the indistinguishability between any subset of the photons and a crucial parameter determining the property of multi-photon interference. Recent efforts have successfully accomplished this complex task by utilising reconfigurable multi-port interfrometers \cite{Brod:2019-63602:PRL, Pont:2022-31033:PRX}. However, the protocol was developed for M-PSSs with a specific form of density matrix. How to characterise multi-photon indistinguishability of M-PSSs with an arbitrary form of density matrix remains an open question. 

Beyond the quantification of indistinguishability, the measurement of the full density matrix can provide comprehensive information about the state, including the collective photon phase.
%Beyond that, an even more complex task is to measure the full density matrix of the M-PSSs, which contains all the information of the states.} 
This is typically done by a process called quantum state tomography, where the density matrix of the input quantum state is reconstructed after a series of projection measurements~\cite{Altepeter:2005-105:AAMOP, Lvovsky:2009-299:RMP, Toninelli:2019-67:ADOP}.
%whose observed quantities are related to the elements of the density matrix~\cite{Altepeter:2005-105:AAMOP, Lvovsky:2009-299:RMP, Toninelli:2019-67:ADOP}. 
To fully reconstruct the density matrix, the number of distinct measurements which are in the form of multi-photon correlations should exceed the number of free parameters in the density matrix that increases exponentially with the number of photons~\cite{Teo:2021-103021:NJP}. To satisfy this requirement, conventional tomography approaches are based on free-space setups or integrated circuits that are reconfigured multiple times~\cite{James:2001-52312:PRA, Shadbolt:2012-45:NPHOT, Pont:2022-31033:PRX}, yet the reconfiguration can be a source of experimental inaccuracies and also 
make the characterization time-consuming for larger numbers of photons.
%and complex 
%The reconfiguration process also introduces more sources of noise. 
On the other hand, static tomography approaches have been suggested~\cite{DAriano:2002-1:PLA, Allahverdyan:2004-120402:PRL, DAriano:2004-165:EPL, Titchener:2016-4079:OL, Banchi:2018-250402:PRL} and realized experimentally~\cite{Titchener:2018-19:NPJQI, Wang:2018-1104:SCI}, where the measurements at the outputs of a fixed photonic circuit enable the full state reconstruction. 
However, such methods have been developed for general states without taking into account the specific structure of PSSs. It remained an outstanding question of how to perform optimal characterization of PSSs with the minimum number of measurements, high robustness to fabrication inaccuracy, and measurement noise while using the most compact and practical photonic circuit design.

%their performance of QST for MSSs is unknown.
% can be further optimized. 
% Therefore, a systematic study on the density matrix form of split states and their QST is missing. The platform optimized for QST of multi-photon split states is still missing.  

In this work, we formulate the general properties of
%perform a systematic investigation on the structure of 
the spatial density matrix structure for the PSSs without any assumptions. Then, we present a scalable approach for single-shot complete state measurement with a static integrated photonic circuit, without a need for reconfigurability.
% Specifically, we propose a multiport coupled waveguide array (CWA) which is segmented into multiple sections along the propagation direction with adjustable local phase changes between adjacent sections. 
% All waveguides are near-neighbour coupled with a same coupling coefficient. 
% Such a CWA configuration represents a new architecture of photonic neural network (NN), where the weights and biases are controlled by the propagation constants of the waveguides and the local phase changes, respectively. This property allows us to numerically train the NN for different missions with optimized performance such as the QST of multi-photon split state as we demonstrated in this paper. 
Specifically, we first theoretically derive the number of free parameters and the structure of the reduced spatial density matrix as a function of the number of photons. Furthermore, we obtain the number of free real and imaginary parts, in which the imaginary values of the density matrix are associated with the multi-photon collective phases. 
% The reduced number of free parameters of MSSs when compared with the general multi-photon states allows us to realise QST of them by a quantum circuit with a simpler configuration. Specifically, 
To realize the state tomography, we propose a multiport coupled waveguide array that is segmented into multiple sections along the propagation direction with the specially introduced local phase shifts between adjacent sections. Such a configuration effectively represents a photonic neural network (NN), where the waveguide coupling and local phase shifts function as the weight and bias, respectively. By optimizing the photonic circuit, we identify the configurations allowing for the most efficient tomography of two-, three- and four-photon split states with reduced sensitivity to measurement noise and fabrication deviations.
%is reduced.
%Specifically, the sensitivity to measurement noise is reduced, corresponding to the lowest values of the condition numbers of 2.2631, 3.9053 and 16.3464, respectively, which are smaller than previous results using other platforms \cite{Miranowicz:2014-62123:PRA, Titchener:2018-19:NPJQI}. 
Different from previous reconfigurable platforms which require multiple measurements with an exponential increase in the number of photons, the proposed scheme can realize the tomography in a single shot without reconfigurability. When compared with previous static approaches developed for general states, the performance is better and the complexity of the photonic circuit is reduced.
%CWA (i.e. the number of required waveguides) is reduced due to a lower number of free parameters for MSSs. 
This makes the proposed scheme scalable to larger photon numbers.

The paper is organized as follows. 
In Sec.~\ref{sec:states}, we formulate the general structure of the reduced spatial density matrix for PSSs after tracing out the internal spectral degree of freedom and determine the numbers of independent real- and complex-valued coefficients as a function of the number of photons. 
In the following Sec.~\ref{sec:circuit}, we introduce a circuit design based on coupled waveguides, representing a linear artificial neural network, and describe its application for split-state tomography.
Then, in Sec.~\ref{sec:results}, we present the circuit designs for two-, three- and four-photon split states, optimized for accurate state reconstruction in presence of measurement noise or fabrication imperfections. Finally, we present conclusions and outlook in Sec.~\ref{sec:conclusion}

%----------------------------------------------
\section{Multi-photon split states and the spatial density matrix} \label{sec:states}

\begin{figure*}[t!]
\centering\includegraphics[width=1\textwidth]{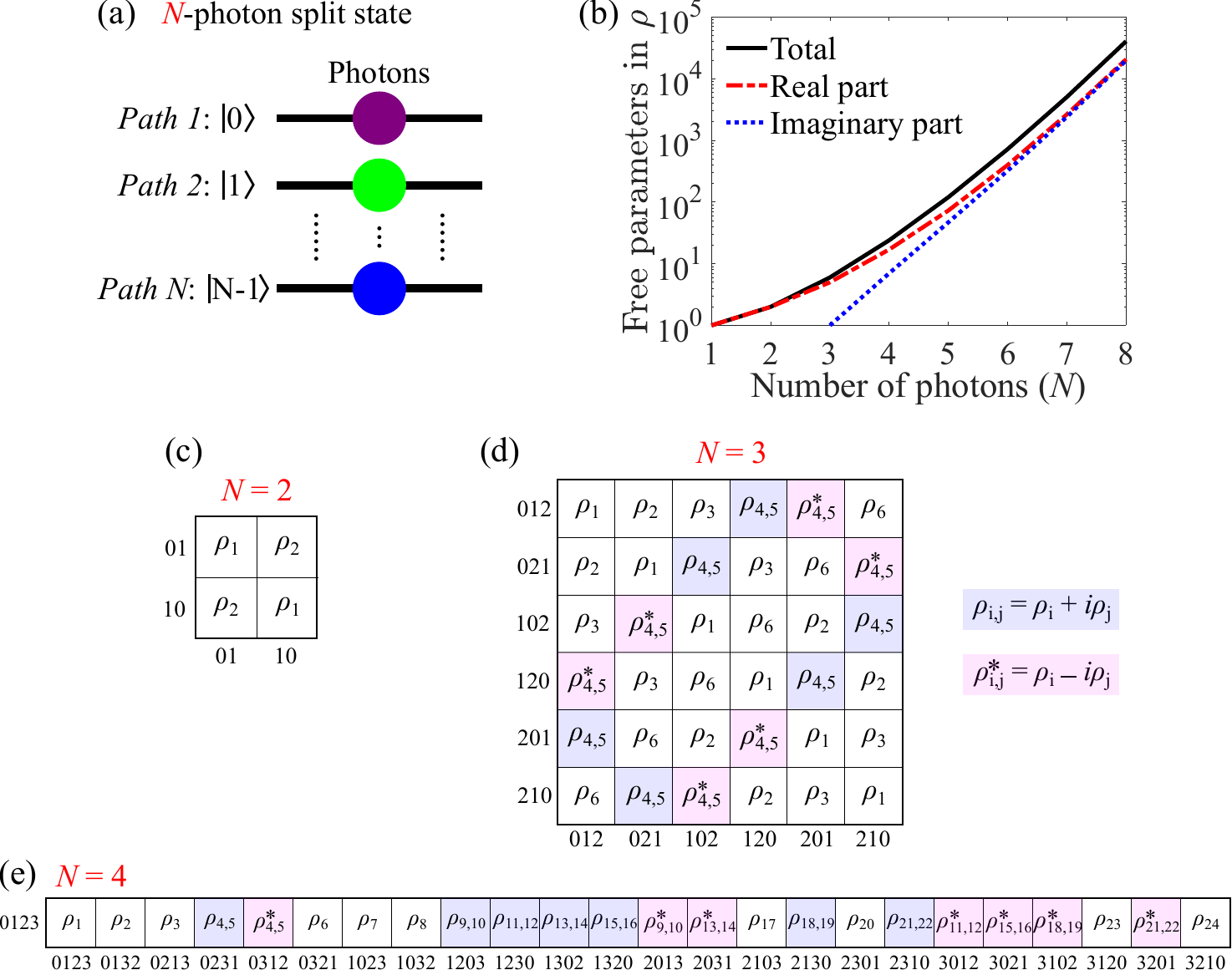}
\caption{\label{fig:split}
(a)~Schematic of the $N$-photon split state (PSS) where each photon is located in a different spatial mode or path.
%All paths are uncoupled, i.e. orthogonal. 
(b)~Number of real, imaginary, and total independent parameters in the reduced spatial density matrix of N-PSS as a function of the number of photons. 
(c,d)~The structure of the reduced spatial density matrices for (b)~two- and (c)~three-PSSs. The elements without or with backgrounds are purely real or complex-valued, respectively. Note that only the nonzero elements of the full density matrix (dimension: $N^N \times N^N$) are shown. (e) First row of the nonzero elements of the four-PSS density matrix. Other rows will have the same elements in different orders.
%(see Supplementary Fig.~\ref{figS1}).
}
\end{figure*}

%We consider the setups using commonly available single-photon click detectors. The coincidences of signals from several detectors then provide a measure of the photon correlations. While the photons might have different internal structure, such as frequency spectra, the conventional detectors only register the arrival time within specific time bins. Such detection and correlation measurements in photonic circuits do not distinguish the photons \textcolor{red}{by their spectral/temporal properties}~\cite{Shadbolt:2012-45:NPHOT, Shchesnovich:2014-22333:PRA}. Also, experiments can realize multi-photon interference that does not explicitly depend on the internal structure~\cite{Menssen:2017-153603:PRL, Titchener:2018-19:NPJQI, Wang:2018-1104:SCI, Jones:2020-123603:PRL}, and such transformations are mathematically described by a unitary operator that mixes different input ports but does not depend on frequency or other internal photon characteristics~\cite{Shchesnovich:2015-13844:PRA, Shchesnovich:2018-33805:PRA}. \textcolor{red}{In addition, the single-photon click detectors cannot resolve the number of photons arrived on the detector. }

We define a multi-photon split state formed by $N$ photons 
%(labeled by $p_0$, $p_1$, $\cdots$, $p_{N-1}$) 
where each photon is located on a different spatial path with orthogonal states labeled by $|0\rangle$, $|1\rangle$, $\cdots$, $|N-1\rangle$, as shown in Fig.~\ref{fig:split}(a). The frequency dependent wavefunction of such a PSS can be expressed as
%~\cite{Titchener:2018-19:NPJQI} 
%(see complete derivation in Appendix~\ref{appendix:elements}):
\begin{widetext}
\begin{equation} \label{eq:wavefun}
|\Psi\rangle = \int d\omega_0 d\omega_1 \cdots d\omega_{N-1} \psi(\omega_0,\omega_1, \cdots, \omega_{N-1}) \hat{a}_{0}^\dag(\omega_0) \hat{a}_{1}^\dag(\omega_1) \cdots \hat{a}_{N-1}^\dag(\omega_{N-1})|0\rangle,
%|\Psi\rangle = \sum_{m_0,m_1, \cdots, m_{N-1}} \int d\omega_0 d\omega_1 \cdots d\omega_{N-1} \psi_{m_0,m_1, \cdots, m_{N-1}}(\omega_0,\omega_1, \cdots, \omega_{N-1}) \hat{a}_{m_0}^\dag(\omega_0) \hat{a}_{m_1}^\dag(\omega_1) \cdots \hat{a}_{m_{N-1}}^\dag(\omega_{N-1})|0\rangle,
\end{equation}
\end{widetext}
where 
%$m_i (i=0, 1, \cdots, N-1)$ is the spatial mode where photon $p_i$ locates and 
$\psi(\omega_0,\omega_1, \cdots, \omega_{N-1})$ is the joint spatial and spectral distribution of $N$ photons. 
%which can be decomposed into the following form for the split state.
%
%\begin{equation}
%\psi(\omega_0,\omega_1, \cdots, \omega_{N-1}) = \phi_{0}(\omega_0)\phi_{1}(\omega_1)\cdots\phi_{{N-1}}(\omega_{N-1}).
%\end{equation}
%
%\end{widetext}
%Here $\phi_{i}(\omega) (i=0, 1, \cdots, N-1)$ represents the spectrum of photon $p_i$. 
Note that the $|0\rangle$ in Eq.~(\ref{eq:wavefun}) stands for the photon vacuum state instead of the spatial mode of the first path mentioned earlier.

We consider the setups using commonly available single-photon click detectors. The coincidences of signals from several detectors then provide a measure of the photon correlations. While the photons might have different internal structure, such as frequency spectra, the conventional detectors only register the arrival time within specific time bins. Such detection and correlation measurements in photonic circuits do not distinguish the photons by their spectral properties~\cite{Shadbolt:2012-45:NPHOT, Shchesnovich:2014-22333:PRA}. Also, experiments can realize multi-photon interference that does not explicitly depend on the internal structure of photons~\cite{Menssen:2017-153603:PRL, Titchener:2018-19:NPJQI, Wang:2018-1104:SCI, Jones:2020-123603:PRL}, and such transformations are mathematically described by a unitary operator that mixes different input ports but does not depend on frequency spectra of photons~\cite{Shchesnovich:2015-13844:PRA, Shchesnovich:2018-33805:PRA}. In addition, the single-photon click detectors cannot resolve the number of photons that arrived on the detector.

For the case where the experimental detectors do not distinguish the photons by their spectrum, 
%and do not resolve the number of photons, 
a PSS can be characterized by a reduced density matrix, where the internal spectrum degree of freedom of the photons is traced out via integration. In the reduced spatial density matrix, which has a dimension of $N^N \times N^N$, each element is determined by~\cite{Titchener:2018-19:NPJQI}:
\begin{widetext}
\begin{equation} \label{eq:rhoReduced}
  \rho_{s'_0,s'_1, \cdots s'_{N-1}; s_0, s_1, \cdots s_{N-1}} = {\rm Tr} \left(\hat{\rho} \, \hat{O}_{s'_0, s'_1, \cdots s'_{N-1}; s_0, s_1, \cdots s_{N-1}}\right) \,,
%  \left(\hat{\rho} \, \ket{s'}\bra{s} \right),
%  =
%Tr \left(\hat{\rho} \, \hat{O}_{(s | s') }\right) ,
%  \rho_{(s_1,s_2...s_N | s_1',s_2'...s_N')} =
%   {\rm Tr} \left(\hat{\rho} \, \hat{O}^{\pm}_{(s_{p_1},s_{p_2}...s_{p_N} | s_{p_1}',s_{p_2}'...s_{p_N}')} \right),
\end{equation}
where $\hat{\rho}=|\Psi\rangle\langle\Psi|$ is the full density matrix, and the $N$-photon density matrix projection operator is defined as
\begin{equation} \label{eq:Operator}
\begin{aligned}
   \hat{O}_{s'_0, s'_1, \cdots s'_{N-1}; s_0, s_1, \cdots s_{N-1}} = \frac{1}{N!} \int d\omega_0 d\omega_1 \cdots d\omega_{N-1} 
   &{\hat{a}_{s'_0}^\dag(\omega_0) \hat{a}_{s'_1}^\dag(\omega_1) \cdots \hat{a}_{s'_{N-1}}^\dag(\omega_{N-1})|0\rangle}\\
   &{\langle0| \hat{a}_{s_0}(\omega_0) \hat{a}_{s_1}(\omega_1) \cdots \hat{a}_{s_{N-1}}(\omega_{N-1})}.
   %\ket{s'_{0}}\bra{s_{0}}\otimes \ket{s'_{1}}\bra{s_{1}}  \otimes ... \ket{s'_{N-1}}\bra{s_{N-1}} .
\end{aligned}
\end{equation}
\end{widetext}
%
%
%\begin{equation}
%    \hat{O}_{(s_{p_1},s_{p_2}...s_{p_N} | s_{p_1}',s_{p_2}'...s_{p_N}')}= \ket{s_{p_1}}\bra{s'_{p_1}}\otimes \ket{s_{p_2}}\bra{s'_{p_2}}  \otimes ... \ket{s_{p_N}}\bra{s'_{p_N}} .
%\end{equation}
%
%Here $s_{p_1} \in (0,K-1)$ denotes the port numbers of the photon labeled $p_1$, and so on.

%$\ket{s},\ket{s'}\in (0,K-1)$ are the labels for the $K$ different ports or waveguide numbers.

%According to \cite{Titchener:2018-19:NPJQI}, the number of free parameters in the density matrix of a state with $N$ indistinguishable photons and $M_{in}$ input ports is $(M_{in}^2+N-1)!/[N!(M_{in}^2-1)!]$. 
For a split state, the nonzero density matrix elements can only be associated with indices $(s'_0, s'_1, \cdots, s'_{N-1})$ and $(s_0, s_1, \cdots, s_{N-1})$ that are permutations in the set $(0, 1, \cdots, N-1)$ without repetitions. We also note that the reduced density matrix is invariant under the simultaneous exchange of indices $s'_{i} \leftrightarrow s'_{j}$ and $s_{i} \leftrightarrow s_{j}$ for arbitrary $i$ and $j$, since the photons are indistinguishable after the internal spectrum degree of freedom is traced out. Using this property, we can map all the elements to just the first row of the density matrix with elements $\rho_{0, 1, \cdots, N-1,s_0s_1 \cdots s_{N-1} }$, where $(s'_0, s'_1, \cdots, s'_{N-1}) = (0, 1, \cdots, {N-1})$.
%
%\begin{equation}
%\begin{aligned}
%\rho_{ 01 \cdots N-1,s_0s_1 \cdots s_{N-1} } =\frac{1}{N!}I_{0,{s_0}}I_{1,{s_1}} \cdots I_{{N-1},{s_{N-1}}}.
%\end{aligned}
%\end{equation}
%
%Here $I_{i,j}$ are the spectral overlaps between photons $p_i$ and $p_j$.
%
%\begin{equation}
%    I_{i,j}=\langle\phi_{i}|\phi_{j}\rangle=\int d\omega\phi_{i}^\ast(\omega)\phi_{j}(\omega),
%\end{equation}
%
%and they satisfy the normalisation $I_{i,i}=1$.
%since the photons are indistinguishable within the reduced density matrix
%
%We label the single-photon state where the photon is in the $j$-th port as $|j-1\rangle$. 
%Then, the number of independent parameters in the density matrix can be determined by the distinct projection operators in the form \cite{Titchener:2018-19:NPJQI}
%
%\begin{equation} \label{eq:operator}
%\hat{O} = |0\rangle \langle s_0| \otimes |1\rangle \langle s_1| \otimes \cdots \otimes |N-1\rangle \langle s_{N-1}| \,,
%\end{equation}
%
%where $(s_0, s_1, \cdots, s_{N-1})$ are permutations in the set $(0, 1, \cdots, N-1)$ without repetitions. 
Therefore, the number of nonzero and independent elements of the spatial density matrix is the number of permutations of $(s_0, s_1, \cdots, s_{N-1})$ in the set $(0, 1, \cdots, N-1)$ without repetition, which is $N!$.
% Therefore, the number of free parameters will be equal to the  number of possible ways of arranging $N$ photons into $N$ ports, that is $N!$. 
%as we discussion
%, which allows us to perform the QST of MSSs with a less complicated quantum circuit in this work.

Within this formulation, we can associate the appearance of collective multi-photon phase~\cite{Shchesnovich:2018-33805:PRA} with the presence of complex-valued density matrix elements. 
%Specifically, the independent real and imaginary parts of the density matrix elements are related to the nonzero distinct $(\hat{O}+\hat{O}_{H.C.})/2$ and $(\hat{O}-\hat{O}_{H.C.})/2$, respectively, where $H.C.$ stands for Hermitian conjugate. 
%
%Among the $N!$ distinct elements, let us denote by $A_N$ the number of cases where $\hat{O}=\hat{O}_{H.C.}$ such that the related imaginary parts are zero. 
Since the ordering of $|0\rangle \langle s_0|, |1\rangle \langle s_1|, \cdots, |N-1\rangle \langle s_{N-1}|$ doesn't affect the values of density matrix elements, we have 
\begin{equation} \label{eq:rho_symm}
  \begin{split}
  \rho_{0, 1, \cdots N-1; s_0, s_1, \cdots s_{N-1}} & = \rho_{ q_0, q_1, \cdots q_{N-1}; 0, 1, \cdots N-1} \\
  & = \rho_{ 0, 1, \cdots N-1; q_0, q_1, \cdots q_{N-1}}^\ast  ,
  \end{split}
\end{equation}
where $(s_0, s_1, \cdots s_{N-1})$ is reordered into $(0, 1, \cdots N-1)$ and $(q_0, q_1, \cdots q_{N-1})$ is the new order from $(0, 1, \cdots N-1)$ after the same permutation operation. When $(q_0, q_1, \cdots q_{N-1}) \equiv (s_0, s_1, \cdots s_{N-1})$, then it follows from Eq.~(\ref{eq:rho_symm}) that $\rho_{ 0, 1, \cdots, N-1; s_0, s_1, \cdots, s_{N-1}} = \rho_{ 0, 1, \cdots, N-1; s_0, s_1, \cdots, s_{N-1}}^\ast$, i.e. this element is real-valued. Let us denote the number of such cases by $A_N$. The remaining $N!-A_N$ elements will have complex values and include $(N!-A_N)/2$ complex-conjugate pairs. Correspondingly, the number of independent real and imaginary parts of the density matrix are $(N!+A_N)/2$ and $(N!-A_N)/2$, respectively, resulting in a total of $N!$ real-valued free parameters. We prove that $A_N$ satisfies the recurrence relation $A_N = A_{N-1} + (N-1)A_{N-2}$ for $N \geq 3$, with $A_1=1$ and $A_2=2$, see Appendix~\ref{appendix:parameters} for the derivation. 

We show in Fig.~\ref{fig:split}(b) the number of total, real, and imaginary independent parameters of the split-state density matrix as a function of the number of photons. For $N=(2,3,4)$, there are $(2, 6, 24)$ total, $(2, 5, 17)$ real, and $(0, 1, 7)$ imaginary parameters. 
%After that, they all follow an exponential increase with respect to the number of photons. 
%
The corresponding structures of the reduced density matrices are presented in Figs.~\ref{fig:split}(c-e).
%, including the nonzero free parameters and the related structures of the density matrices. 
We label the independent real and imaginary coefficients with single sequential indices. The Figs.~\ref{fig:split}(c) and \ref{fig:split}(d) illustrate that one row contains all the different elements, and we show just the first row for four-PSS in Figs.~\ref{fig:split}(e) to save space. 
%Note that 
%they are the reduced version of the full density matrices and 
%all the unlabelled elements are zero for the split states. 
Notably, the imaginary parts can only appear for three or more photons ($N \ge 3$), in agreement with the properties of collective multi-photon phase~\cite{Menssen:2017-153603:PRL, Shchesnovich:2018-33805:PRA, Jones:2020-123603:PRL}.

We emphasise that the number of independent density matrix parameters for split states ($N!$) is much smaller than that for the general states~\cite{Titchener:2018-19:NPJQI}.
%, especially for high photon numbers
This allows for the efficient tomography of split states by taking into account their structure, as we discuss in the following.

%These reduced density matrices can be reconstructed from the multi-photon correlation measurements, as will be shown later. 

%-----------------------------------
\section{Integrated circuit for split-state measurements} \label{sec:circuit}

\begin{figure*}[ht!]
\centering\includegraphics[width=1\textwidth]{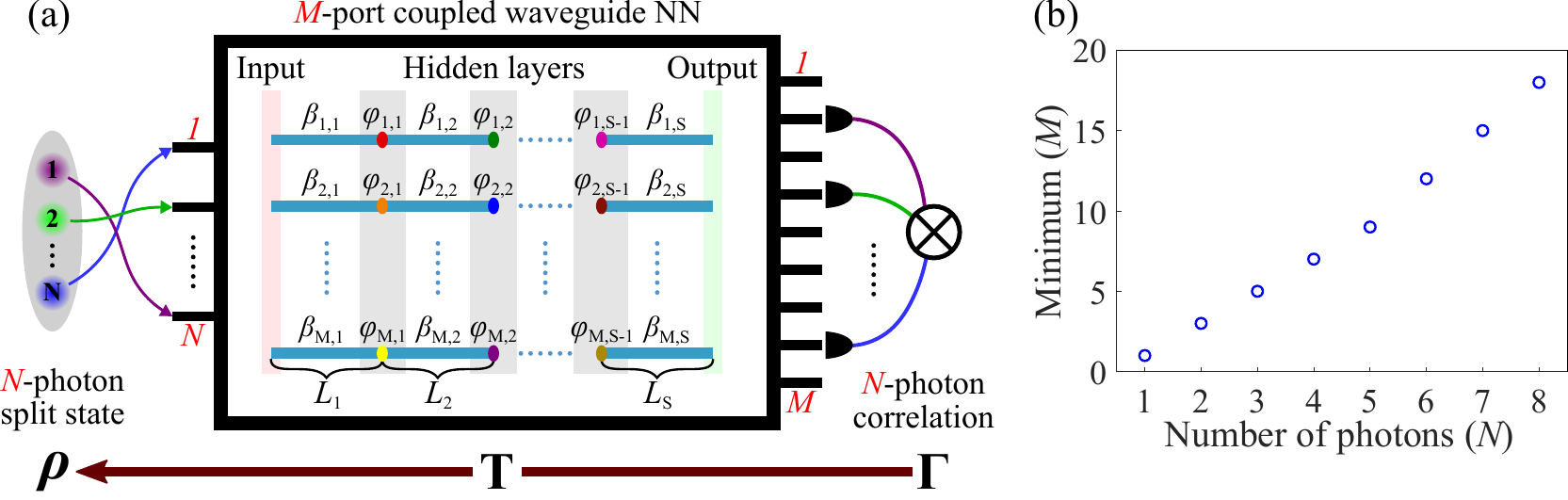}
\caption{\label{fig:circuit}
(a) Schematic of the proposed $M$-port coupled waveguide neural network for $N$-photon split state tomography, where the output $N$-photon correlations enable the reconstruction of the input density matrix. (b) Required number of waveguides for different photon numbers. }
\end{figure*}

We now analyse how the characterization of split states can be performed by adopting a static tomography approach~\cite{DAriano:2002-1:PLA, Allahverdyan:2004-120402:PRL, DAriano:2004-165:EPL, Titchener:2016-4079:OL, Banchi:2018-250402:PRL, Titchener:2018-19:NPJQI, Wang:2018-1104:SCI}. 
We consider an $N$-input-$M$-output waveguide circuit where the $N$ input photons interfere, and the $N$-photon correlations are measured between different combinations of the $M$ output ports, as schematically shown in Fig.~\ref{fig:circuit}(a). 
To realise the state tomography without reconfigurability, the number of different $N$-photon correlation measurements at the output should exceed the number of unknown density matrix elements, that is 
\begin{equation} \label{eq:Mmin}
    \frac{M!}{N!(M-N)!} \geq N! \,.
\end{equation}
The required minimum number of waveguides $M$ grows linearly up to five-photon states, $M_{\rm min} =2N-1$ for $N \leq 5$. At larger photon numbers, we obtain an exact quadratic fitting as $M_{\rm min} = \lceil 0.139 N^2+1.174N-0.387 \rceil$ for $N \leq 30$, see Fig.~\ref{fig:circuit}(b).
%and Fig.~\ref{figS2}.
We confirm the quadratic scalability at high photon numbers using Stirling's approximation, which provides an asymptotic estimate $M>(N/e)^2 \simeq 0.135 N^2$ for $N \gg 1$.
%,  For infinite number of photons, we prove that the condition of $M$ is $M>(N/e)^2$ using Stirling's approximation, which maintains the quadratic trend and thus indicates practical scalability of our scheme. 

Since no structure tunability is required, there is a large design freedom of the input-to-output transformation based on integrated waveguide circuits. We demonstrate the general approach of split-state tomography for the realization based on arrays of straight coupled waveguides~\cite{Christodoulides:2003-817:NAT, Zhou:2018-3020:OE, Skryabin:2021-26058:OE} where the undesirable bending losses are absent, since higher transmission is critical for the observation of multi-photon interference~\cite{Garcia-Patron:2019-169:QUA}.
Furthermore, the continuous inter-waveguide coupling along the propagation direction can make the circuit
%and has no bends in waveguides, making it a 
% and less lossy structure when 
more compact compared with the commonly used scheme of cascaded Mach–Zehnder interferometers \cite{Crespi:2013-545:NPHOT, Clements:2016-1460:OPT}, while allowing for the realization of various quantum logic operations~\cite{Lahini:2018-2:NPJQI}. 

The proposed waveguide circuit is sketched inside the central frame in 
Fig.~\ref{fig:circuit}(a).
%shows the schematic of the proposed CWA for the QST of MSSs.
It consists of $M$ waveguides, which optical modes are coupled to the nearest-neighbours with the constant coupling coefficient $\kappa$. The waveguides are segmented into $S$ sections with lengths $(L_1, L_2, \cdots, L_S)$. 
%In each of the sections, the propagation constants of optical modes can be detuned by the values $\beta_{i,j}$. 
We consider the presence of tailored phase shifts $\varphi_{i,j}$ at the interfaces between adjacent sections, noting that such localized shifts were demonstrated experimentally~\cite{Szameit:2008-181109:APL} and their incorporation 
%in a similar architecture based on multi-core fibers 
was predicted to allow 
%can be used to realize 
arbitrary unitary transformations \cite{Zhou:2018-3020:OE, Saygin:2020-10501:PRL}.
%, there is a local phase shift that can be adjusted.
In addition, the design based on straight and identical waveguides  ensures that there is no mismatch in the photon propagation lengths and dispersion, and the circuit is expected to better preserve the degree of photon indistinguishability during their interference compared to cascaded interferometers~\cite{Bell:2019-35646:OE}.
By design, the losses are expected to be similar for all waveguides, and we assume in the following that the output multi-photon correlations are above the noise level~\cite{Garcia-Patron:2019-169:QUA}, which can be satisfied experimentally for at least $N=5$ photons~\cite{Bell:2019-35646:OE}.
%dispersion and beamsplitter ratio, which will introduce less photon distinguishability due to imperfection of the multi-port interferometer~\cite{Bell:2019-35646:OE}. 
Essentially, the configuration in Fig.~\ref{fig:circuit}(a) represents a linear artificial neural-network~\cite{Steinbrecher:2019-60:NPJQI} with $S-1$ hidden layers, where each hidden layer has $M$ neurons. The waveguide couplings in each section function as the weights and the local phase shifts play similar roles to the bias. 

The overall linear system transformation of the multi-photon state, provided that the losses are negligible, can be determined by a classical or one-photon unitary transfer matrix
%of the \textit{M}-port CWA-NN is:
\begin{equation}
   \textbf{U}=\textbf{W}_S~\textbf{B}_{S-1}~\textbf{W}_{S-1}~\cdots~\textbf{B}_2~\textbf{W}_2~\textbf{B}_1~\textbf{W}_1 \,.
\end{equation}
Here $\textbf{W}_j=\exp(i\textbf{C} L_j)$, calculated through the matrix exponent, is the weight matrix of the layer $j$, where
%between layers $j+1$ and layer $j$ with 
the coupling matrix elements are $\textbf{C}_{n,m} = \kappa \delta_{n,m\pm1}$.
%defined by
%\begin{equation}
%\textbf{C}_j=\begin{pmatrix}
%\beta_{1,j} & \kappa & 0 & \cdots & 0\\
%\kappa & \beta_{2,j} & \kappa & \cdots & 0\\
%\vdots & \ddots & \ddots & \ddots & \vdots\\
%0 & \cdots & \kappa & \beta_{M-1,j} & \kappa\\
%0 & \cdots & 0 & \kappa & \beta_{M,j}
%\end{pmatrix}.
%\end{equation}
The bias matrix acting on the $j$-th hidden layer is  $\textbf{B}_j = \exp(i \Phi_j)$, where the exponent is applied element-wise to the phase shift matrix $\Phi_j = {\rm diag}(\varphi_{1,j}, \varphi_{2,j}, \ldots , \varphi_{M,j} )$.
%\begin{equation}
%\Phi _j=\begin{pmatrix}
%\varphi_{1,j} & 0 & \cdots & 0 & 0\\
%0 & \varphi_{2,j} & 0 & \cdots & 0\\
%\vdots & \ddots & \ddots & \ddots & \vdots\\
%0 & \cdots & 0 & \varphi_{M-1,j} & 0\\
%0 & 0 & \cdots & 0 & \varphi_{M,j}
%\end{pmatrix}.
%\end{equation}
%
The
%is system transformation is an 
$M \times M$ unitary matrix $ \textbf{U}$ can be flexibly tuned by varying the lengths of sections
%, propagation constant detunings 
and local phase shifts. 
%Indeed, it was found that a similar architecture based on multi-core fibers can realize an arbitrary unitary transformation \cite{Zhou:2018-3020:OE}.
%, suggesting that the structure can be efficiently optimized 

%, which is essential in the integrated quantum network \cite{Crespi:2013-545:NPHOT, Clements:2016-1460:OPT}.

%\begin{figure}[ht!]
%\centering\includegraphics[width=1\columnwidth]{fig2.pdf}
%\caption{\label{fig:circuit}
%(a) Schematic of the proposed $M$-port coupled waveguide neural network for $N$-photon split state tomography, %where the output $N$-photon correlations enable the reconstruction of the input density matrix. (b) Required %number of waveguides for different photon numbers. }
%\end{figure}

%-------------------------
%\section{Optimize the CWA-NN for QST of MSSs}
%\section{Optimize the CWA-NN for QST of MSSs}

For an $N$-photon split state, where the photons are coupled to specific $N$ ports at the input, its transformation is governed by \textit{N-in-M-out} matrix $\textbf{U}_r$, which contains $N$ columns of $\textbf{U}$ corresponding to the selected inputs.
%we use the unitary matrix 
%Since we consider an 
%For the QST of a \textit{N}-photon split state, we choose \textit{N} ports at the input layer and construct an \textit{N-in-M-out} NN with a transformation $\textbf{U}_r$ which is the related $N$ columns of $\textbf{U}$. 
Then, we use a standard procedure~\cite{Titchener:2018-19:NPJQI, Wang:2018-1104:SCI} to calculate the output $N$-photon correlations, which values can be represented by a vector $\vec{\Gamma}$ of the length equal to different combinations $M! / [N! (M-N)!]$. The correlations can be expressed through the  independent elements of the input density matrix arranged in a vector $\vec{\rho}_{\rm free}$ of length $N!$,
%This reduced transformation correlates the probability amplitudes of the output $N$-photon correlations (number: $C_M^N$) with the free parameters of the input density matrix (number: $N!$), resulting in a relation
\begin{equation} \label{eq:Gamma}
    \vec{\Gamma} = \textbf{T}~\vec{\rho}_{\rm free} \,,
\end{equation}
where the matrix $\textbf{T}$ is determined by $\textbf{U}_r$ and the structure of the split state density matrix. 

Finally, we can reconstruct the input density matrix based on the measured $N$-photon correlations as 
\begin{equation} \label{eq:rhoInverse}
 \vec{\rho}_{\rm free}=\textbf{T}^+ \, \vec{\Gamma} \,,
\end{equation} 
where $\textbf{T}^+$ is the pseudoinverse of $\textbf{T}$.

It is essential to optimize the integrated circuit to reduce the reconstruction sensitivity to noise and measurement errors.
Mathematically, this is achieved by minimizing the condition number of the matrix $\textbf{T}$, defined as the ratio of the transformation’s maximum and minimum singular values $\sigma_{max}(\textbf{T})/\sigma_{min}(\textbf{T})$ \cite{Press:2007:NumericalRecipes}. 
For our structure design, we perform numerical optimization of the waveguide section lengths and the local phase shifts
%to minimize the condition number of the reconstruction matrix $\textbf{T}$ which is defined as $\sigma_{max}(\textbf{T})/\sigma_{min}(\textbf{T})$, the ratio of the transformation’s maximum and minimum singular values \cite{Press:2007:NumericalRecipes}. 
%This condition number quantifies the amplification of measurement errors and noise during the reconstruction process. 
%Our numerical algorithm is 
based on the Nelder-Mead simplex direct search algorithm realized using the \textit{fminsearch} function in Matlab.

%------------------------------------------------
\section{Results and discussions} \label{sec:results}

We performed extensive simulations of coupled-waveguide neural networks and found that tomography of split states with the photon number at least up to four can be  efficiently performed in structures where all sections have the same length, all waveguides have the same propagation constants and thus zero detunings, and all the near-neighbour waveguide couplings are equal to each other. These conditions make the photonic circuit design and fabrication simpler, where all the waveguides have the same widths and the spacings between them are identical.
%In this work, we utilize this CWA-NN for QST of MSSs. For this purpose, we simplify the CWA-NN by letting all sections have the same length, all waveguides have the same propagation constants and thus zero detunings, and the coupling coefficient be normalized to one. These simplifications facilitate the easy and high-quality fabrication. 
We show in the following that optimization of the local phase shifts and the total waveguide length ($L$) allows us to reach low condition numbers, corresponding to low sensitivity to noise during the reconstruction. To simplify the notations, we consider the scaling of the waveguide length in the units of $\kappa^{-1}$, such that the coupling coefficient is normalized to one.

\begin{figure}[ht!]
\centering\includegraphics[width=1\columnwidth]{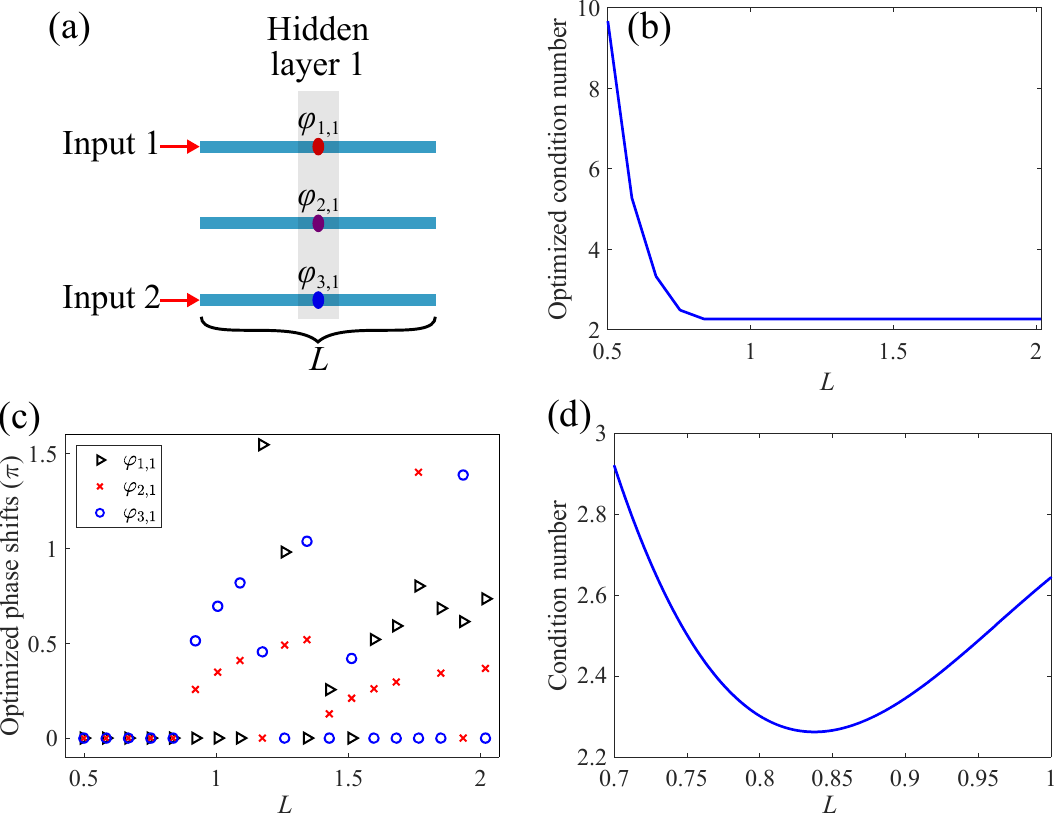}
\caption{\label{fig:2photon}
Optimized circuit for two-PSS tomography.
%Training the CWA-NN for QST of two-photon split state. 
(a)~The structure of three coupled waveguides 
%of the \textit{2-in-3-out} CWA-NN 
with one hidden layer.
%for QST of two-photon split state. 
(b,c)~Optimized (b)~condition number and (c)~phase shifts at the hidden layer as a function of the total waveguide length. 
%(c) The related optimized phase shifts after training. 
(d)~Condition number for a structure without a hidden layer vs. the waveguide length.
%s as a function of the total waveguide length near $L=0.84$ when there is no hidden layer. 
}
\end{figure}

We first analyze the tomography of two-PSS. We choose the minimum required number of $M=3$ waveguides according to Eq.~(\ref{eq:Mmin}) and Fig.~\ref{fig:circuit}(b), select the first and third waveguides as the input ports, and consider a circuit structure with one hidden layer ($S=2$) as sketched in Fig.~\ref{fig:2photon}(a).
%We show in Fig.~\ref{fig:2photon}(a) the optimized condition numbers as a function of total waveguide length for the tomography of two-photon split state with one hidden layer, i.e. $S=2$. The number of waveguides is $M=3$ and we choose the first and third waveguides as the input ports. 
We perform the optimization for different waveguide lengths and show the best condition number values in Fig.~\ref{fig:2photon}(b).
One can see that the condition number reaches a minimum value of $\simeq 2.3$ when the waveguide length is longer than 0.84. This optimized condition number is smaller than the previously reported values for tomography of general  two-photon states~\cite{Titchener:2016-4079:OL, Titchener:2018-19:NPJQI}.
% Note that this minimum value is the theoretical limit for the two-photon split state [?]. 
The corresponding optimized phase shifts at the hidden layer are shown in Fig.~\ref{fig:2photon}(c), where we assign zero to one of the phases since the global phase does not affect the output correlations. 
Interestingly, all the three phase shifts are zero for the waveguide length shorter than 0.84, which effectively corresponds to the absence of hidden layer.
%is equivalent to a NN without hidden layer and phase shifts. This will make the experimental implementation much easier. 
For longer waveguides, the minimum value of condition number 
%After this critical length of 0.84, 
is achieved for circuits with an optimal hidden layer.
%NN can still reach the minimum value of condition number with hidden layers. 
For comparison, Fig.~\ref{fig:2photon}(d) shows the condition number for a structure without a hidden layer.
We see that the circuit can allow for optimal performance over a broad range of structure lengths, offering more flexibility in integrating with other photonic components.
%as a function of the waveguide length near $L=0.84$ when there is no hidden layer. 
%We see see that the NN can maintain a small condition number in a large length range, indicating a high tolerance to the fabrication errors on the length.

\begin{figure}[ht!]
\centering\includegraphics[width=1\columnwidth]{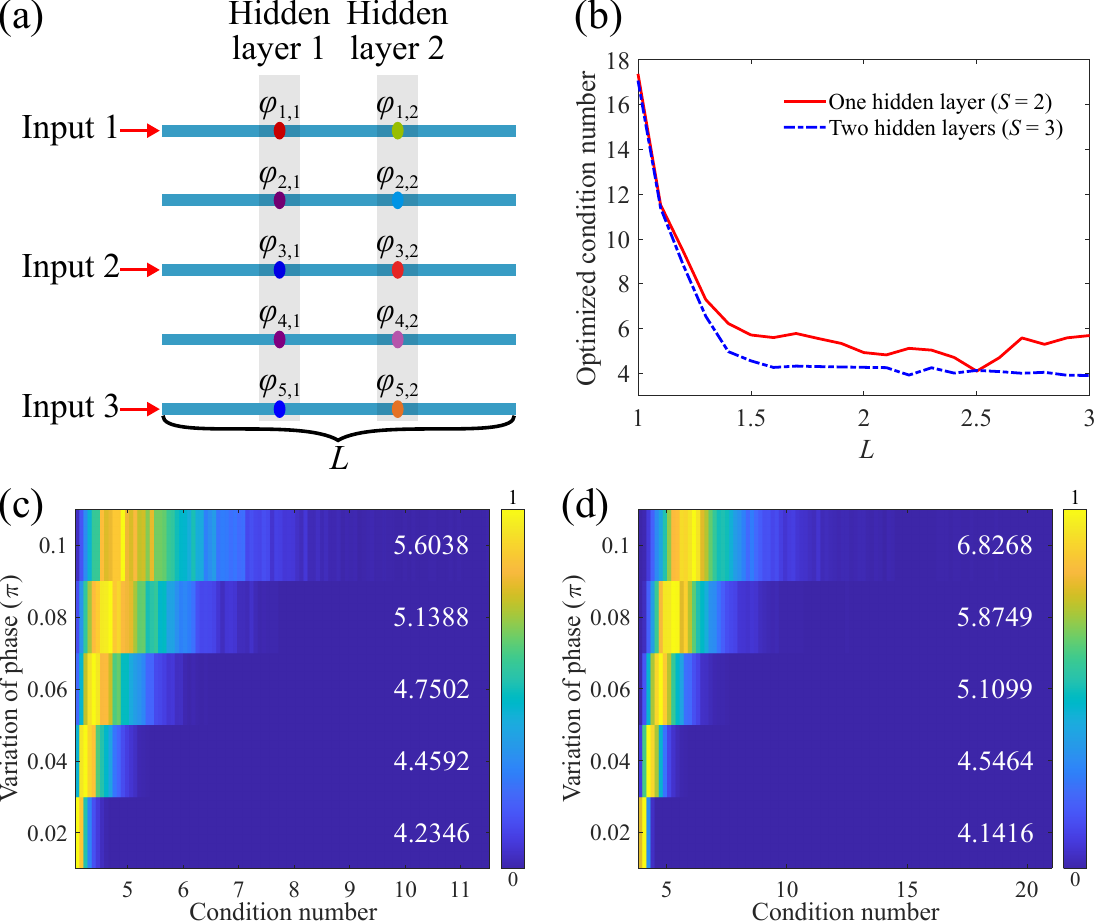}
\caption{\label{fig:3photon}
%Training the CWA-NN for QST of
Tomography of three-PSS.
%three-photon split state. 
(a)~The structure of the \textit{3-in-5-out} couped-waveguide network with two hidden layers. (b)~Optimized condition numbers vs. the total waveguide length with one or two hidden layers. (c,d)~Normalized probability density of condition number values for 5000 simulations with random variations of the phase shifts from the optimal values for structures with 
%. The simulations are done for 
(c) one hidden layer ($S=2$, $L=2.5$) and (d) two hidden layers ($S=3$, $L=3$). The vertical axis represents the magnitude of the random variations and the white-colored numbers are the corresponding averaged condition numbers.
%of the 5000 simulation for each variation magnitude. 
}
\end{figure}

Next, we investigate the three-PSS tomography. Then, we use  Eq.~(\ref{eq:Mmin}) to determine the required number of waveguides as $M=5$ and choose the first, third, and fifth waveguides as the input ports, see an illustration in Fig.~\ref{fig:3photon}(a). We check that without hidden layers, the condition number is very high, which would prevent a state reconstruction.
%, see a discussion in Appendix~\ref{appendix:3nohidden}.
%The next interesting question is that can this CWA-NN enable high-quality QST of higher $N$ without hidden layers. To answer this question, we next study three-photon split state, for which the number of waveguides is $M=5$ and we choose the first, third, and fifth waveguides as the input ports (Fig. \ref{fig:3photon}(a)). Figure S3 shows the condition number as a function of waveguide length when there is no hidden layer and no optimization. The condition number is extremely high, meaning that for higher number of photons, we need to utilize an NN with hidden layers. 
The condition number dependencies on the structure length with the optimized one or two hidden layers are presented in Fig.~\ref{fig:3photon}(b).
%shows the optimized result for three-photon split states with one and two hidden layers.
%Different from the two-photon case, there is no clear limit on the condition number.
Overall, for a certain length, more hidden layers can provide lower condition numbers due to more tuning parameters. The smallest condition numbers are $\simeq 4.1$ at $L=2.5$ and $\simeq 3.9$ at $L=3$ for one and two hidden layers, respectively. These values are much smaller than the ones for general three-photon states~\cite{Wang:2019-41:OPT}. 
% REMOVE GLOBAL PHASE !!!
%The corresponding optimized phase shifts are $\varphi_{j,1}=(4.1620, 5.2446, 5.3291, 5.1349, 3.3876)$ for one hidden layer and $\varphi_{j,1}=(0.0964, 4.3442, 3.9042, 1.5387, 5.1943$, $\varphi_{j,2}=(4.7971, 0.3581, 0.4618, 1.5022, 2.6686$ for two hidden layers.
The corresponding optimized phase shifts are $\varphi_{j,1}=(0,	1.083,	1.167,	0.973,	5.509)$ for one hidden layer and $\varphi_{j,1}=(0,	4.248,	3.808,	1.442,	5.098)$, $\varphi_{j,2}=(0,	1.844,	1.948,	2.988,	4.155)$ for two hidden layers.

We confirm the practicality of the designs by quantifying the tolerance of the optimal structures for three-PSS tomography to variations of the phase shifts due to potential fabrication errors. Figures~\ref{fig:3photon}(c) and~\ref{fig:3photon}(d) show the normalized probability density of the condition number values for random deviations of the phase shifts from the optimal values in different variation ranges. The white-colored numbers are the average condition numbers for different variation magnitudes. At small deviations, the structure with two hidden layers has better performance with the smaller condition number. In case of phase shift variations of $0.04\,\pi$ or larger, the structure with one hidden layer is better. This is because there are fewer phase shifts, and the performance is more robust to their variations. Overall, the condition numbers are smaller than 7, even when the phase shifts vary from the optimized values by a magnitude up to $0.1\pi$. This confirms the high fabrication tolerance of the circuits.
%proposed CWA-NN for QST of three-photon states.

\begin{figure}[ht!]
\centering\includegraphics[width=1\columnwidth]{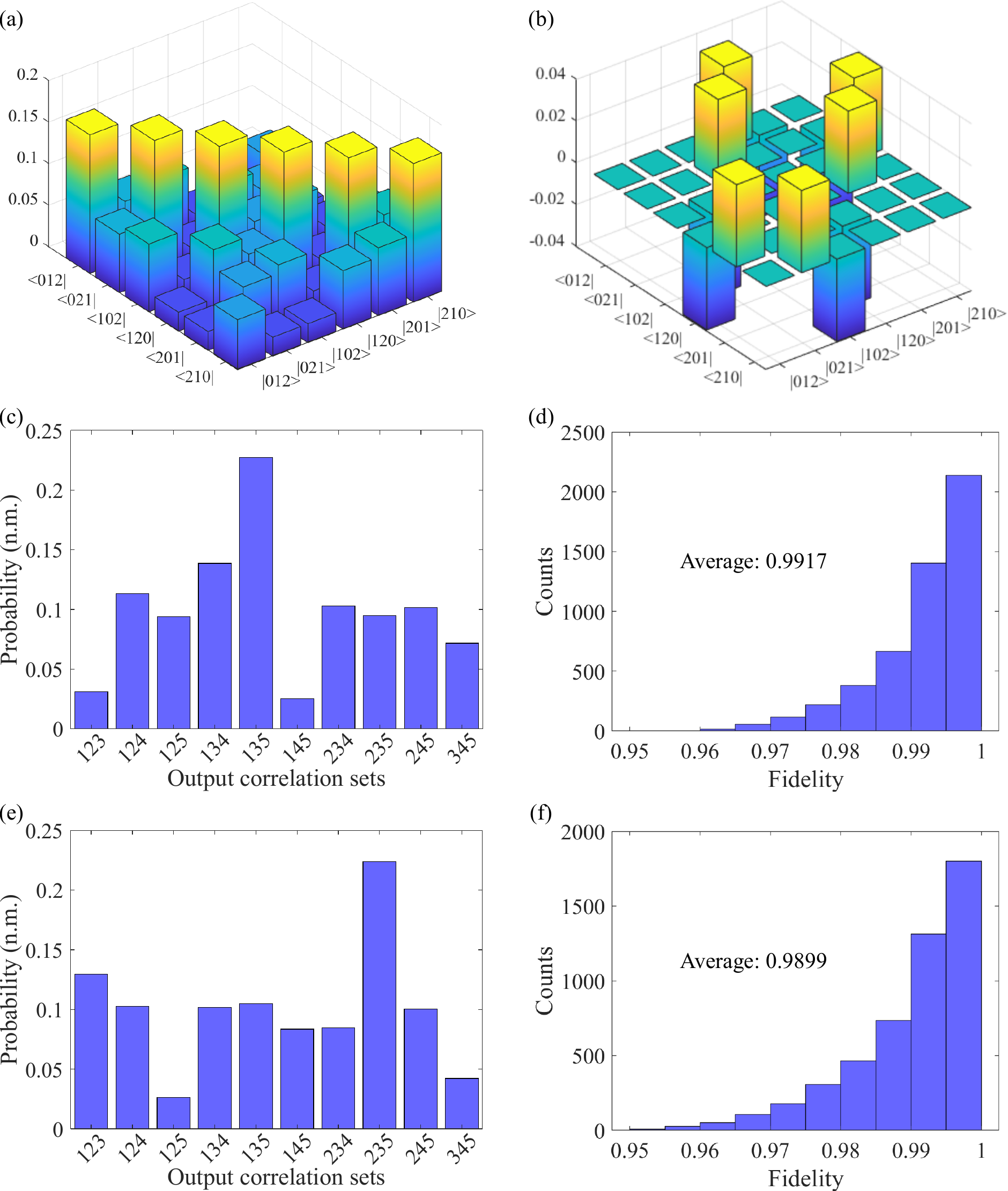}
\caption{\label{fig:3correlations}
Three-photon reconstruction fidelity in presence of measurement noise.
(a)~Real and (b)~imaginary parts of the density matrix for an input three-photon split state with a collective phase of $-\pi/3$. (c,e)~The predicted three-photon correlation probabilities after passing through the optimized waveguide network. (d) The statistical distribution of the density matrix reconstruction fidelity for 5000 simulations when a Gaussian noise with a standard deviation of $5\%$ is added to the output correlations. 
Results correspond to the structures with (c,d)~one and (e,f)~two hidden layers.
%(e) The predicted three-photon correlation probabilities after passing through the designed CWA-NN with two hidden layers and (f) the counts distribution of the density matrix reconstruction fidelity for 5000 simulations when a random Gaussian noise with a standard deviation of $5\%$ is added to the predicted correlation probabilities.
}
\end{figure}
Next, we demonstrate numerically the density matrix reconstruction of three-PSS.
%apply the optimized designs to reconstruct two specific three-photon split states. 
%For the first state, 
%\textcolor{red}{
%
As an example, we consider the three-PSS composed of photons with uncorrelated frequency spectra, defined by the wavefunction
%\begin{widetext}
%\begin{equation}
$|\Psi\rangle = \int d\omega_0 d\omega_1 d\omega_2 \phi_{0}(\omega_0)\phi_{1}(\omega_1)\phi_{2}(\omega_2) \hat{a}_0^\dag(\omega_0) \hat{a}_1^\dag(\omega_1) \hat{a}_2^\dag(\omega_2)|0\rangle$,
%\end{equation}
%\end{widetext}
where $\phi_{j}(\omega)$ represents the spectral wavefunction of one photon in the $j$-th spatial path.
%
%As an example, we consider the photons with the pairwise spectral overlaps (labelled $p_0$, $p_1$, and $p_2$) as $\langle\phi_{0}|\phi_{1}\rangle=0.7e^{-i\pi/3}$, $\langle\phi_{1}|\phi_{2}\rangle=0.65$, and $\langle\phi_{2}|\phi_{0}\rangle=0.6$ with a collective phase of $-\pi/3$.
In simulations, we assume the pairwise spectral overlaps as $\langle\phi_{0}|\phi_{1}\rangle=0.7e^{-i\pi/3}$, $\langle\phi_{1}|\phi_{2}\rangle=0.65$, and $\langle\phi_{2}|\phi_{0}\rangle=0.6$, which define all the spatial density matrix elements and the collective photon phase of $-\pi/3$ as formulated in  Appendix~\ref{appendix:elements}.
%\sout{where  $\phi_{i}(\omega)$ represents the spectrum of photon $p_i$. }
The real and imaginary parts of the density matrix of this state are presented in Figs.~\ref{fig:3correlations}(a) and~\ref{fig:3correlations}(b), respectively. Figures~\ref{fig:3correlations}(c) and~\ref{fig:3correlations}(e) show the predicted three-photon correlation probabilities at the output of the optimized circuits
%when inputting this three-photon state into the designed CWA-NN 
with one and two hidden layers, respectively. We see that the correlations are different for each structure.
%CWA-NNs with different transformations. 
Based on the output correlations, one can reconstruct the input density matrix. In order to verify the low sensitivity to the measurement noise, we apply a Gaussian noise to the correlations  and use them to reconstruct the input density matrix. We quantify the quality of the tomography procedure by the fidelity between the reconstructed ($\rho_{rec}$) and the input ($\rho_{th}$) density matrices, defined as ${\rm Tr}(\sqrt{\sqrt{\rho_{th}}\rho_{rec}\sqrt{\rho_{th}}})$. Figures~\ref{fig:3correlations}(d) and~\ref{fig:3correlations}(f) show the  corresponding statistical distributions of the reconstruction fidelity for 5000 simulations when a Gaussian noise with a standard deviation of $5\%$ is added to the predicted correlation probabilities. We find that the fidelity stays above 0.95 for both one- and two- hidden layer structures, with the average values of $\simeq 0.99$.
%$\simeq 0.992$ and $\simeq 0.990$, respectively. 
We also confirm similarly high fidelity for different three-PSSs, including those with a zero collective phase.
%is provided in Appendix Fig. \ref{fig9}. 
These results indicate the high accuracy of the tomographic reconstruction of split states under the presence of measurement noise.

The proposed approach and its high tolerance to fabrication errors and shot noise are also applicable to a larger number of photons. For example, we numerically designed a circuit with $N = 4$, $M = 7$, $S=3$, and~$L=3$ for the tomography of four-PSS with an optimized condition number of 16.3464, for the hidden-layer phase shifts $\varphi_{j,1}=(0,	1.126,	0.306,	4.331,	4.990,	1.633,	2.419
)$, $\varphi_{j,2}=(0,	1.212,	1.998,	2.246,	0.371,	6.002,	0.894)$ 
%(the phase shifts are provided in Appendix Table~\ref{table:4shifts}). 

%After numerically obtained the optimized phase shifts, the CWA-NN 
The designed circuits can be realized experimentally based on different integrated photonic platforms with established fabrication techniques. The localized phase shifts in coupled segmented waveguides were achieved through 
%can be experimentally demonstrated based on different integrated photonic platforms such as 
fs-laser writing in silica~\cite{Szameit:2008-181109:APL} and phase control using voids inside the waveguide was shown in silicon photonic circuits~\cite{Wang:2019-3547:NCOM}.
%waveguides or integrated silicon photonic circuits. 
%For the fs-laser writing waveguides, different local phase shifts can be realized by tuning the laser power, which results in different propagation constants of the waveguides.
% \cite{Mennea:2018-1087:OPT}.
%For silicon waveguides, 
%The local phase shifts can be realized by 
%local resonators 
%via waveguide segmentation~\cite{Szameit:2008-181109:APL} or voids inside the waveguide\cite{Wang:2019-3547:NCOM}.
%, which have been utilized to realize on-chip metasurface \cite{Wang:2019-3547:NCOM}. 

%Beyond the characterization the multi-photon split state, the circuits proposed in this work can also be tailored for other integrated quantum applications, such as measuring the indistinguishability and entanglement of identical photons \cite{Tichy:2013-225:RAR}, and performing photonic boson sampling \cite{Broome:2013-794:SCI, Tillmann:2013-540:NPHOT, Crespi:2013-545:NPHOT}. Specifically, it is possible to extend the algorithm in this work to consider waveguides with losses, which always exist in practically demonstrated waveguides and play an essential role in realizing the quantum advantage \cite{Garcia-Patron:2019-169:QUA}.

\section{Conclusion} \label{sec:conclusion}

To conclude, we have formulated the general structure of spatial density matrix for multi-photon split states, which are an important resource for various quantum applications and whose resource-efficient characterization is a sought-after capability. We then proposed a coupled waveguide array forming a photonic neural network for the quantum tomography of such states with low sensitivity to noise and high tolerance to fabrication errors. The state measurement can be performed using a static photonic circuit and this approach is scalable to high photon numbers. 
%and to other subsets of multi-photon states such as the NOON state where all photons are located in the same mode.} 

We anticipate that the proposed platform, enabling simple and robust characterization of such commonly used quantum states, will stimulate further developments and applications of quantum optical circuits. In particular, since our scheme does not require reconfigurability, it is especially suitable for integration with on-chip superconducting nanowire single-photon detectors operating at cryogenic temperatures to facilitate plug-and-play split-state measurements.
Furthermore, the theoretical methodology can be extended to the split states where photons are separated not only in spatial ports but also in other degrees of freedom, such as polarization and orbital angular momentum.

% \section{Backmatter}

% Backmatter sections should be listed in the order Funding/Acknowledgment/Disclosures/Data Availability Statement/Supplemental Document section. An example of backmatter with each of these sections included is shown below.

%\begin{backmatter}
\begin{acknowledgments}
%\bmsection{Funding}
This work is supported by the Australian Research Council (DP190100277).
%
%\bmsection{Acknowledgments}
Authors acknowledge useful discussions with Sidi Lu, Kai Wang, and Alexander Szameit.
%
%\bmsection{Disclosures}
%The authors declare no conflicts of interest.
%
%
%\bmsection{Data Availability Statement}
%
%Data underlying the results presented in this paper are not publicly available at this time but may be obtained from the authors upon reasonable request.
The simulation data underlying the results
presented in this paper may be obtained from the authors upon reasonable request.
\end{acknowledgments}
%A Data Availability Statement (DAS) will be required for all submissions beginning 1 March 2020. The DAS should be an unnumbered separate section titled ``Data Availability'' that
%immediately follows the Disclosures section. See \href{https://www.osapublishing.org/submit/review/data-availability-policy.cfm}{OSA's Data Availability Statement policy page} for more information.

%\end{backmatter}

%\newpage
%\section{Appendix}
\appendix

%%\newcounter{MainEquations}
%%\setcounter{MainEquations}{\value{equation}}
%%\renewcommand{\theequation}{A$\the\numexpr\value{equation}-\the\numexpr\value{MainEquations}$}

%\renewcommand{\thesection}{\Alph{section}}
%\titlespacing*{\section}{0pc}{0pt}{0pt}

%\begin{figure}[ht!]
%\centering\includegraphics[width=0.85\columnwidth]{figS1.pdf}
%\caption{\label{figS1}
%Structure of the reduced density matrix for four-photon split state. }
%\end{figure}

%\begin{figure}[ht!]
%\centering\includegraphics{figS2.pdf}
%\caption{\label{figS2}
%Quadratic fit of minimum required number of waveguides up to 30 photons. }
%\end{figure}

%\begin{figure}[ht!]
%\centering\includegraphics{figS3.pdf}
%\caption{\label{figS3}
%Optimized condition number for QST of three-photon split state when there is no hidden layer. }
%\end{figure}
\section{Number of real and imaginary free parameters in density matrices of split states} \label{appendix:parameters}

%As stated in the main text, t
As derived above, the number of nonzero elements in the reduced density matrix of $N$-photon split state equals to the distinct projection operators in the form of Eq.~(\ref{eq:Operator}), and it is $N!$.
%
%\sout{
%\begin{equation}
%\hat{O} = |0\rangle \langle s_0| \otimes |1\rangle \langle s_1| \otimes \cdots \otimes |N-1\rangle \langle s_{N-1}|,
%\end{equation}
%}%
%\textcolor{red}{
%\begin{equation} \label{eq:Operator}
%\begin{aligned}
 %  \hat{O} = \frac{1}{N!} \int d\omega_0 d\omega_1 \cdots d\omega_{N-1} 
  % &{\hat{a}_{0}^\dag(\omega_0) \hat{a}_{1}^\dag(\omega_1) \cdots \hat{a}_{N-1}^\dag(\omega_{N-1})|0\rangle}\\
   %&{\langle0| \hat{a}_{s_0}(\omega_0) \hat{a}_{s_1}(\omega_1) \cdots \hat{a}_{s_{N-1}}(\omega_{N-1})},
   %\ket{s'_{0}}\bra{s_{0}}\otimes \ket{s'_{1}}\bra{s_{1}}  \otimes ... \ket{s'_{N-1}}\bra{s_{N-1}} .
%\end{aligned}
%\end{equation}
%}
%
%where $(s_0, s_1, \cdots, s_{N-1})$ are permutations in the set $(0, 1, \cdots, N-1)$ without repetition, resulting in $N!$ distinct $\hat{O}$. 
The numbers of real and imaginary parts are determined by the numbers of nonzero distinct $(\hat{O}+\hat{O}_{H.C.})/2$ and $(\hat{O}-\hat{O}_{H.C.})/2$, respectively, where $H.C.$ stands for Hermitian conjugate~\cite{Titchener:2018-19:NPJQI}. To perform the counting, we define by $A_N$ the number of cases when $\hat{O}=\hat{O}_{H.C.}$, such that the related imaginary parts are zero. 
%Accordingly, the numbers of free real and imaginary parts are $(N!+A_N)/2$ and $(N!-A_N)/2$, respectively. 
Next, we derive the recurrence relation for $A_N$ as a function of $N$. Let us consider the value of $s_{N-1}$. 
When $s_{N-1}=N-1$, the number of cases where $\hat{O}=\hat{O}_{H.C.}$ is $A_{N-1}$. When $s_{N-1} = \tilde{n}$ for $\tilde{n} = (0, 1, \ldots, N-2)$, the condition $\hat{O}=\hat{O}_{H.C.}$ can only be satisfied when $s_{\tilde{n}}=N-1$. In this case, the number of cases where $\hat{O}=\hat{O}_{H.C.}$ becomes $A_{N-2}$. 
Since $\tilde{n}$ can take $N-1$ values, the total number is $(N-1)A_{N-2}$. 
Therefore, we obtain the relation 
\begin{equation}
    A_N = A_{N-1} + (N-1)A_{N-2} \quad {\rm for} \quad N\geq3, 
\end{equation}
and the values for one- and two-photon states are $A_1=1$ and $A_2=2$.
%We confirmed this result with 

%\newpage
%----------------------------------------------------------------
\begin{widetext}
\section{The spatial split-state density matrix for photons with uncorrelated frequency spectra}  \label{appendix:elements}

Whereas our approach is applicable to arbitrary multi-photon split states, here we discuss an example of states composed of photons with uncorrelated frequency spectra.
Specifically, we consider a pure $N$-photon state
\begin{equation}
|\Psi\rangle = 
\int \,d\omega_0\,d\omega_1\,\cdots, d\omega_{N-1} 
\psi(\omega_0,\omega_1, \ldots \omega_{N-1}) \hat{a}_{0}^\dag(\omega_0) \hat{a}_{1}^\dag(\omega_1) \cdots \hat{a}_{N-1}^\dag(\omega_{N-1})|0\rangle \, 
\end{equation}
with the frequency-dependent wavefunction featuring no correlations between the individual spectra of photons,
\begin{equation}
\psi(\omega_0,\omega_1, \ldots, \omega_{N-1}) 
= \phi_{0}(\omega_0)\phi_{1}(\omega_1) \cdots \phi_{N-1}(\omega_{N-1}) \,.
\end{equation}
Here $\phi_{j}(\omega_j)$ is an individual spectral wavefunction of the photon coupled to a spatial mode number $j$.
We calculate the $N!$ nonzero elements of the first row of the reduced density matrix for the $N$-photon split state as
\begin{equation}
\begin{split}
&{ \rho_{ 0,1, \cdots, N-1; s_0, s_1, \cdots, s_{N-1} } = {\rm Tr} \left(\hat{\rho} \, \hat{O}_{0, 1, \cdots, N-1; s_0, s_1, \cdots, s_{N-1}}\right)} \\
&\quad { = \frac{1}{N!}\int d\omega_0 d\omega_1 \cdots d\omega_{N-1} \phi_{0}^\ast(\omega_0)\phi_{1}^\ast(\omega_1)\cdots\phi_{{N-1}}^\ast(\omega_{N-1}) }
{\phi_{{s_0}}(\omega_0)\phi_{{s_1}}(\omega_1)\cdots\phi_{{s_{N-1}}}(\omega_{N-1}) } \\
&\quad { =\frac{1}{N!}I_{0,{s_0}}I_{1,{s_1}} \cdots I_{{N-1},{s_{N-1}}} } \,,
\end{split}
\end{equation}
where $(s_0, s_1, \cdots, s_{N-1})$ are permutations in the set $(0, 1, \cdots, N-1)$ without repetition and
we define the spectral overlaps between different photon pairs as 
\begin{equation}
I_{i,j}=\langle\phi_{i}|\phi_{j}\rangle=\int d\omega\phi_{i}^\ast(\omega)\phi_{j}(\omega) \,,
\end{equation}
with the normalization $I_{j,j}=1$.

For an $N=3$ photon case, we have
\begin{equation}
\begin{aligned}
    &{ \rho_{0,1,2;0,1,2} 
    =\frac{1}{6} } \,, \,\,
    { \rho_{0,1,2;0,2,1} 
    =\frac{1}{6}|I_{1,2}|^2 } \,, \,\,
    { \rho_{0,1,2;1,0,2} 
    =\frac{1}{6}|I_{0,1}|^2 } \,, \,\,
    { \rho_{0,1,2;2,1,0} 
    =\frac{1}{6}|I_{2,0}|^2 } \,, \\
    &{ \rho_{0,1,2;1,2,0} 
    =\frac{1}{6}I_{0,1} I_{1,2} I_{2,0} } \,,  \,\,
    {\rho_{0,1,2;2,0,1} 
    =\frac{1}{6}I_{0,1}^\ast I_{1,2}^\ast I_{2,0}^\ast} .  \quad
\end{aligned}
\end{equation}
We see that the elements $\rho_{0,1,2;201} = \rho_{0,1,2;120}^\ast$ are related to the collective phase of three photons. Correspondingly, the six free parameters in Fig.~\ref{fig:split}(d) are
\begin{equation}
\begin{aligned}
&{ \rho_1 = \rho_{0,1,2;0,1,2} } 
 , \quad 
 { \rho_2 = \rho_{0,1,2;0,2,1}  } 
 , \quad 
 { \rho_3 = \rho_{0,1,2;1,0,2}  } \,, \\
&{ \rho_4 = {\rm Re}(\rho_{0,1,2;1,2,0})  } 
 , \quad 
 { \rho_5 = {\rm Im}(\rho_{0,1,2;1,2,0})  } 
 , \quad 
 { \rho_6 = \rho_{0,1,2;2,1,0} } \,.
\end{aligned}
\end{equation}

\end{widetext}

%%%%%%%%%%%%%%%%%%%%%%% References %%%%%%%%%%%%%%%%%%%%%%%%%
\bibliography{art_split_state_tomography}

\end{document}